# Numerical Modeling of Pulse Wave Propagation in a Stenosed Artery using Two-Way Coupled Fluid Structure Interaction (FSI)

Peshala P. T Gamage[1*], Fardin Khalili[1], Hansen A Mansy[1]

[1]University of Central Florida, Department of Mechanical and Aerospace Engineering, Biomedical Acoustic Research Lab, Orlando, FL 32816, USA

## ABSTRACT

As the heart beats, it creates fluctuation in blood pressure leading to a pulse wave that propagates by displacing the arterial wall. These waves travel through the arterial tree and carry information about the medium that they propagate through as well as information of the geometry of the arterial tree. Pulse wave velocity (PWV) can be used as a non-invasive diagnostic tool to study the functioning of cardiovascular system. A stenosis in an artery can dampen the pulse wave leading to changes in the propagating pulse. Hence, PWV analysis can be performed to detect a stenosed region in arteries. This paper presents a numerical study of pulse wave propagation in a stenosed artery by means of two-way coupled fluid structure interaction (FSI). The computational model was validated by the comparison of the simulated PWV results with theoretical values for a healthy artery. Propagation of the pulse waves in the stenosed artery was compared with healthy case using spatiotemporal maps of wall displacements. The analysis for PWV showed significance differences between the healthy and stenosed arteries including damping of propagating waves and generation of high wall displacements downstream the stenosis caused by flow instabilities. This approach can be used to develop patient-specific models that are capable of predicting PWV signatures associated with stenosis changes. The knowledge gained from these models may increase utility of this approach for managing patients at risk of stenosis occurrence.

## 1. INTRODUCTION

Physicians measure the pulse wave velocity (PWV) of patients to measure the arterial stiffness which is an early diagnostic tool of cardiac/arterial diseases [1, 2]. Hence, pulse wave propagation has been a key topic among researches over the years and many studies have been conducted both experimentally [3, 4] and numerically [5, 6].
Some numerical studies have employed fluid-structure interaction (FSI) to simulate the hemodynamic and mechanical characteristics of arteries[5-10]. While most studies have neglected pulse wave propagation [8, 9, 11], a handful have used FSI to study pulse wave propagation using FSI [5-7]. Among these, one study[10] numerically validated the pulse wave velocity by comparison with a theoretically calculated value for a straight artery. Another study [6] concluded that arterial geometry has a strong impact on pulse wave propagation using simulation of pulse wave velocity in straight and curved arteries. A recent study [5] also analysed the effect of geometry and stiffness of aneurysms on PWV and pulse wave propagation using FSI simulations and showed that the presence of an aneurysm can be potentially detected based on the differences in wall displacements and fluid velocity propagation profiles in comparison to a straight artery.

   Development of plaques inside arteries is accompanied with vessel narrowing, known as stenosis, which can increase pressure drop across the stenosed region and cause high blood pressure. In addition, a stenosis affects the wave propagation characteristics since it reflects and dampens the travelling wave. A few studies addressed pulse wave propagation in a stenosed artery [12, 13] and predicted significant damping of the pulse across the stenosis and a strong dependence of the reflection coefficient modulus with the severity of the stenosis[13].

*Corresponding Author: peshala@knights.ucf.edu

The purpose of the current investigation is to study the effect of vascular stenosis on pulse wave velocity and propagation using FSI simulation. First, the simulation is validated by numerically calculating the PWV in a straight healthy artery and comparing it to a theoretical model. Then, the effect of the stenosis on PWV is discussed by analysing the variations of wall displacement using spatiotemporal maps. While PWV is known to be a useful diagnostic tool for coronary and peripheral artery [14] disease, the current results suggest that it may also provide valuable information as a non-invasive tool to detect arterial stenosis.

## 2. MATERIALS AND METHODS

### 2.1 Governing Equations

Blood flow in larger arteries are generally considered as Newtonian fluid [15]. Hence, Newtonian flow is assumed in the current study and incompressible N-S equations are employed which are represented as:

$$\rho_b \frac{\partial U}{\partial t} + \rho_b U.\nabla U - \nabla.\tau = 0 \qquad (1)$$

$$\nabla.U = 0 \qquad (2)$$

$$\tau = -p\delta_{ij} + 2\mu\varepsilon_{ij} \qquad (3)$$

$$\varepsilon_{ij} = \frac{1}{2}(\nabla U + \nabla.U^T) \qquad (4)$$

In above equations $U, p, \tau, \varepsilon_{ij}$ represent the velocity vector, static pressure, shear stress tensor and strain rate respectively. $\delta_{ij}$ is the kronecker delta function. Subscripts $i$ and $j$ denote the cartesian tensor notations. Blood density $\rho_b$ is set to $1050\ kgm^{-3}$ and viscosity $\mu$ was set to $0.00035\ Pa.s$ [16].

In this study, arterial walls are considered as an isotropic linear elastic material and the governing equations are;

$$\rho_w \frac{\partial^2 d_i}{\partial t^2} = \frac{\partial \sigma_{ij}}{\partial x_j} + F_i \qquad (5)$$

$$\sigma = D\epsilon \qquad (6)$$

Where $d_i, F_i, \sigma_{ij}$ represents the components of displacement, body force and stress tensor respectively. Matrix of elastic constants $D$ is determined from the Young's modulus and Poisson's ratio of the material. For the artery wall Young's modulus and Poisson's ratio were set to 5 MPa and 0.499 respectively [16].

### 2.2 Model Geometry

Fig. 1 (a) shows the model geometry used for the current simulation. The fluid domain was extended to avoid the entrance effects. The dimensions of the artery were selected such that they match actual human sizes based on relevant literature [15]. A cross-sectional view of the stenosis is shown in Fig. 1 (b). This shape was modeled according to Eqn. 7 [17].

$$R(x) = R_0(1 - 0.4\ e^{-0.02x^2}) \qquad (7)$$

Where, $R(x)$ is the radius of the artery in the stenosis region and $R_0$ is the radius of the artery in heathy region. In the current model, $R_0$ is 10 mm and wall thickness in healthy region is 1mm.

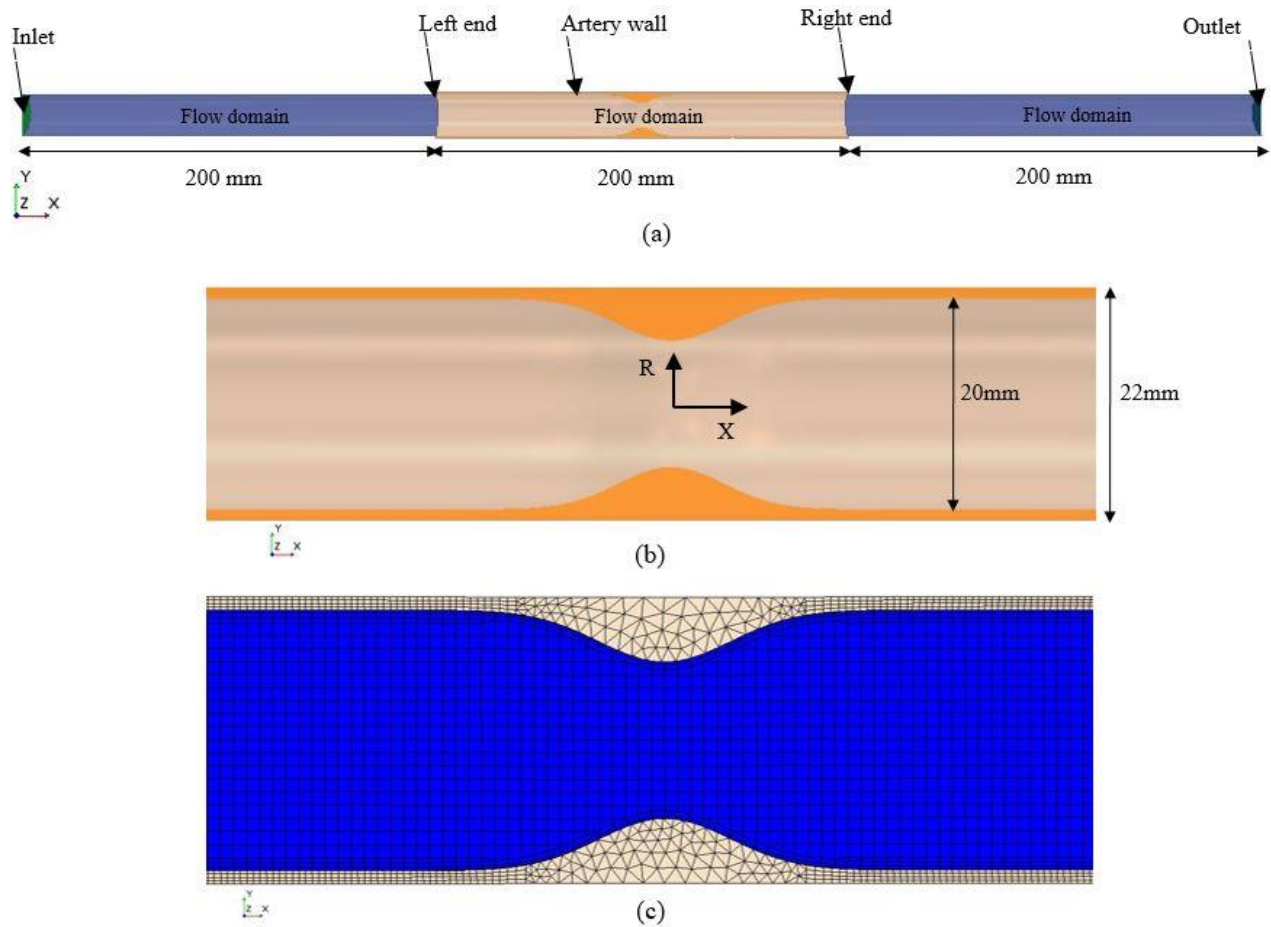

**Fig.** 1  (a) Domain for the simulation (b) Dimensions of the stenosed artery (c) Mesh

## 2.1 Numerical Modeling

*Fluid-Structure Interaction (FSI)*

Two-way coupling interaction in the software package Star CCM+ (STAR-CCM+, CD-adapco, Siemens, Germany) is used in the current study where the fluid pressure affects the displacement of the structure and vice versa. Here, fluid flow and solid displacements were solved using finite volume (FV) and finite element (FE) methods respectively. The data between solid and fluid domain mesh was mapped using a mapped contact interface (MPI) [18] available in the software. Modelling the deformation of the fluid domain of the artery requires redistribution of mesh nodes in response to the calculated displacement in the solid region. This was achieved by employing the dynamic mesh morphing technique [19] where fluid mesh vertices are moved in a way to conform to the solid structure where cells maintain the same neighbors but their shapes can change over time. Second order accuracy was used for both space and temporal discretization.

*Meshing*

Fluid domain was meshed using hexahedron type cells while solid domain was meshed using tetrahedral cells. After a grid independence study [20-22], a total number of 100,135 cells were used in fluid domain while 69,876 cells were used in solid domain. A zoomed view of the cross section of the mesh is shown in Fig. 1 (c).

*Boundary conditions*

A time dependent velocity profile (Fig. 2) and zero static pressure boundary condition were imposed at the inlet and outlet respectively. Non-slip boundary condition was specified on walls. The movement of the left end of the artery was restrained in all direction while the right end was allowed to move only in the axial direction. Artery wall was set free in all degrees of freedom. Mesh morphing feature in Star CCM+ was activated in both domains to simulate the deformations.

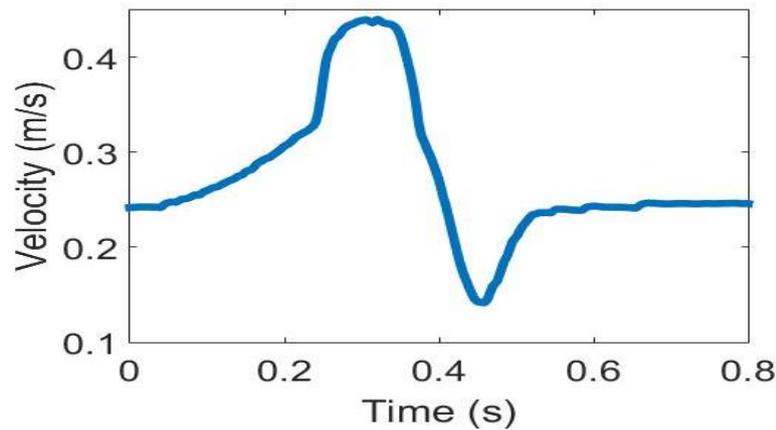

**Fig**. 2. Inlet velocity profile

### 3. RESULTS

**3.1 Validation for Straight Artery**

The pulse wave propagation in a healthy artery (without the stenosis) was simulated and the pulse wave velocity (PWV) value from the simulation was compared with theoretical value calculated based on Kortweg-Moens wave speed equation [23]. In Eqn. 8, *t* and *D* denote the thickness of the artery wall and inner diameter of the artery respectively.

$$PWV = \sqrt{\frac{Et}{\rho_b D}} \quad (8)$$

Two point probes were created on the wall separated by 18 cm in the axial direction. Wall displacement in Y direction was recorded and the pulse wave velocity (PWV) was calculated based on delay between displacement signals at these points. The data was acquired when the simulation was stabilized after completing 3 cycles of the inlet velocity profile. The simulation time step was set to 0.005 seconds. The displacement signals were later interpolated to a time step of 0.0001 seconds and delays were calculated based on cross-correlation [24] between the signals. The delay was found to be 0.015 seconds and PWV was estimated as 15.65 ms$^{-1}$ which was slightly higher than the theoretical PWV (15.43 ms$^{-1}$) calculated using Eqn. 8. The differences (<1.5%) between the calculated values can be due to the, thin wall assumption used in deriving Eqn. 8 [23] and the effect of boundary conditions in the simulation. Furthermore, a spatiotemporal map of Y displacement was plotted along the wall surface (Fig.3) and it was observed that the displacements along the wall propagates in the straight tube with unnoticeable reflections.

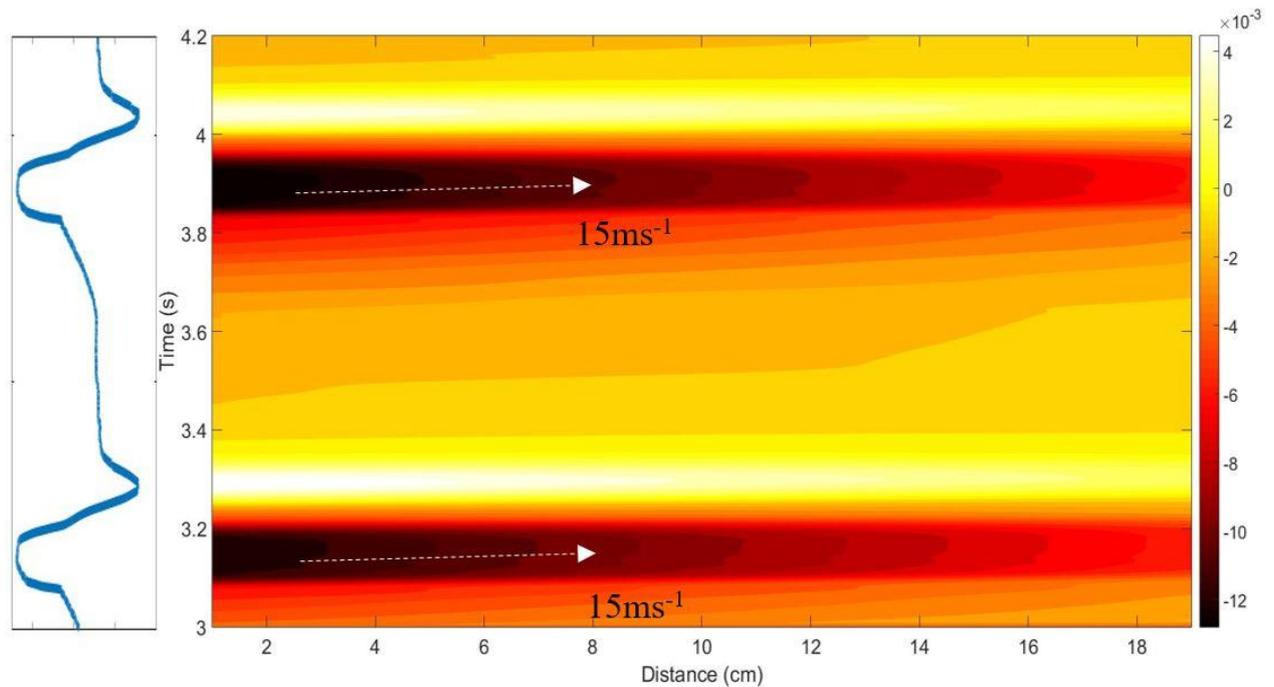

**Fig**. 3  Spatiotemporal map of the displacement (mm) in Y direction on wall, Left column: inlet velocity profile

The high and low value regions of Fig.3 occurs at the peak systolic and diastolic flow rates. The propagation velocity of these high and low displacements along the artery wall can be estimated by the gradient of these regions in the spatiotemporal map [5]. As indicated in Fig. 3, the gradient of these regions was approximately close to previously calculated value using cross correlation.

## 3.2 Pulse Wave Propagation in Stenosed Artery

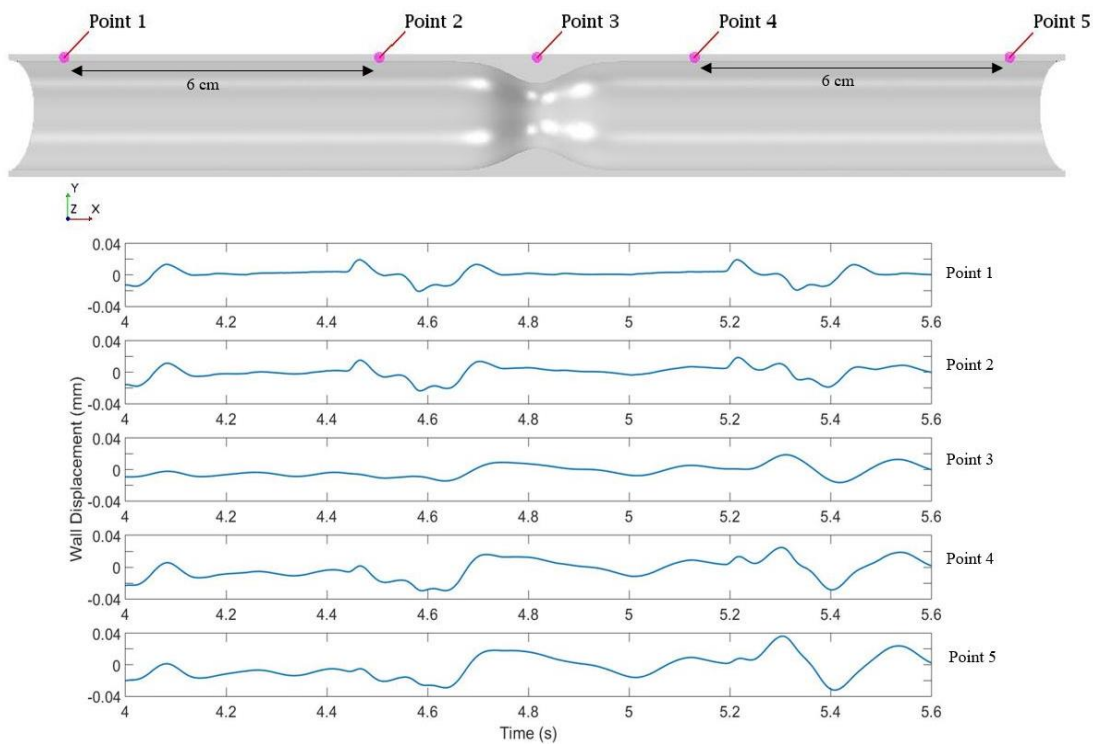

**Fig**. 4  Y-Displacement at different points on artery wall

Wall displacement signals recorded at different point probes on the artery wall is shown in Fig. 4. The amplitude of the displacement on the stenosed region was low which is likely due to the thickening of the wall caused by the stenosis. Furthermore, the amplitude of the signals before the stenosis (acquired at point 1 and 2) was lower than the amplitude of the signals after the stenosed region (acquired at point 4 and 5). To further analyze these signals in frequency domain, fast fourier transformation (FFT) of the signals were plotted.

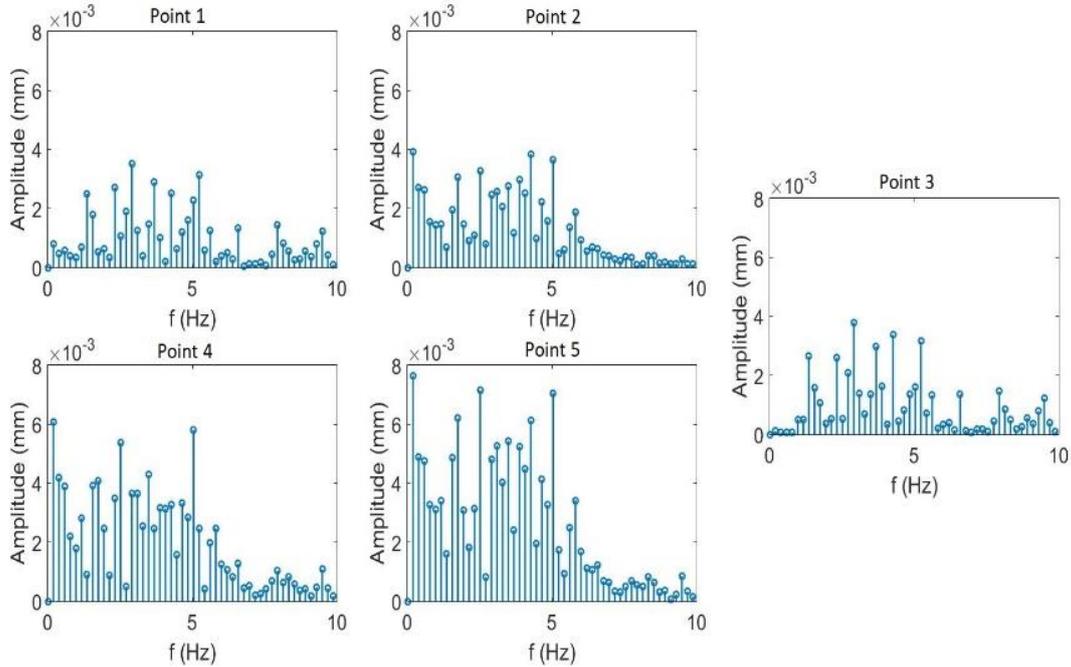

**Fig**. 5  FFT spectra of the recorded Y-displacements on the wall

As shown in Fig. 4 and 5, displacement signals had distinctive features from each other. Also, the spectra at point 2,4,5 had similarities while spectra at point 1,3 were similar to each other. An attempt was given to calculate PWV separately for the regions before and after the stenosis based on the cross-correlation between signals at 1&2 and 4&5. These results are shown in table 1.

**Table 3**  PWV estimate for the stenosed artery based on cross-correlation

| Points | Distance(cm) | Delay(s) | PWV (m/s) |
|--------|--------------|----------|-----------|
| 4&5    | 6            | 0.0036   | 16.66     |
| 1&2    | 6            | 0.0052   | 11.53     |

The PWV before the stenosed region was 24% less than the theoretical value calculated for a heathy artery using Eqn. 8. PWV after the stenosis increased by 44% compared to the PWV before the stenosis. These differences are likely caused by the adverse pressure gradient developed across the stenosis as well as the increased stiffness in the wall caused by the higher thickness in the stenosed region. Significant difference between the calculated PWV and theoretical PWV is expected since Kortweg-Moens equation is derived assuming a constant wall thickness and for a flat flow profile [23]. Also, finding the delay between two signals with different features (such as points 2 and 3 or 3 and 4) using cross-correlation may not provide accurate results. Hence the PWV between these points were not calculated. However, the data for the healthy case suggest that PWV would be comparable to the theoretical value in far upstream and downstream region where the flow is stabilized.

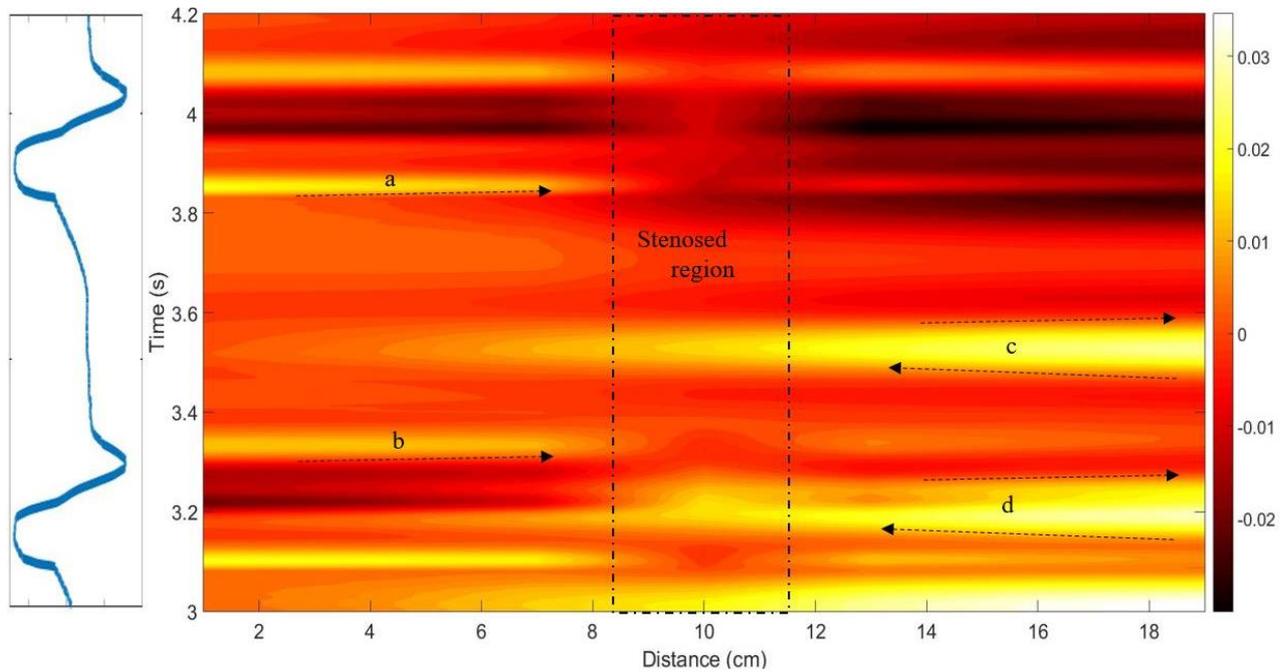

**Fig**. 6 Spatiotemporal map of the displacement (mm) in Y direction on wall for the stenosed artery, Left column: inlet velocity profile

The spatiotemporal map of the Y-displacement (Fig. 6) had more variations compared to the healthy case (Fig. 3). Some propagating waves (a and b) were dampened at the stenosis, while some waves (c and d) seems to increase in magnitude downstream the stenosis. The propagation directions of the displacements in these regions are shown by arrows based on their gradients. While the displacements in regions a and b propagates in the forward direction, they seem to follow a cyclic pattern with respect to the inlet velocity profile. The damping of these waves possibly caused by the increased wall thickness in the stenosed region. In region c and d, some high displacements tended to propagate forward while some propagated backward based on the gradients in spatiotemporal map. This is likely due to the high wall displacements behind the stenosis possibly caused by the breakage of the velocity jet generated at the constriction and reattachment of the flow in the downstream where part of the waves propagates backward while some propagate forward. Fig. 7 shows the velocity and wall displacement on a cross-section of the artery at different times during the cardiac cycle. The time-dependent inlet velocity is show in the right column. These results clearly show the velocity instabilities and breakage of the jet and recirculation zones downstream the stenosis.

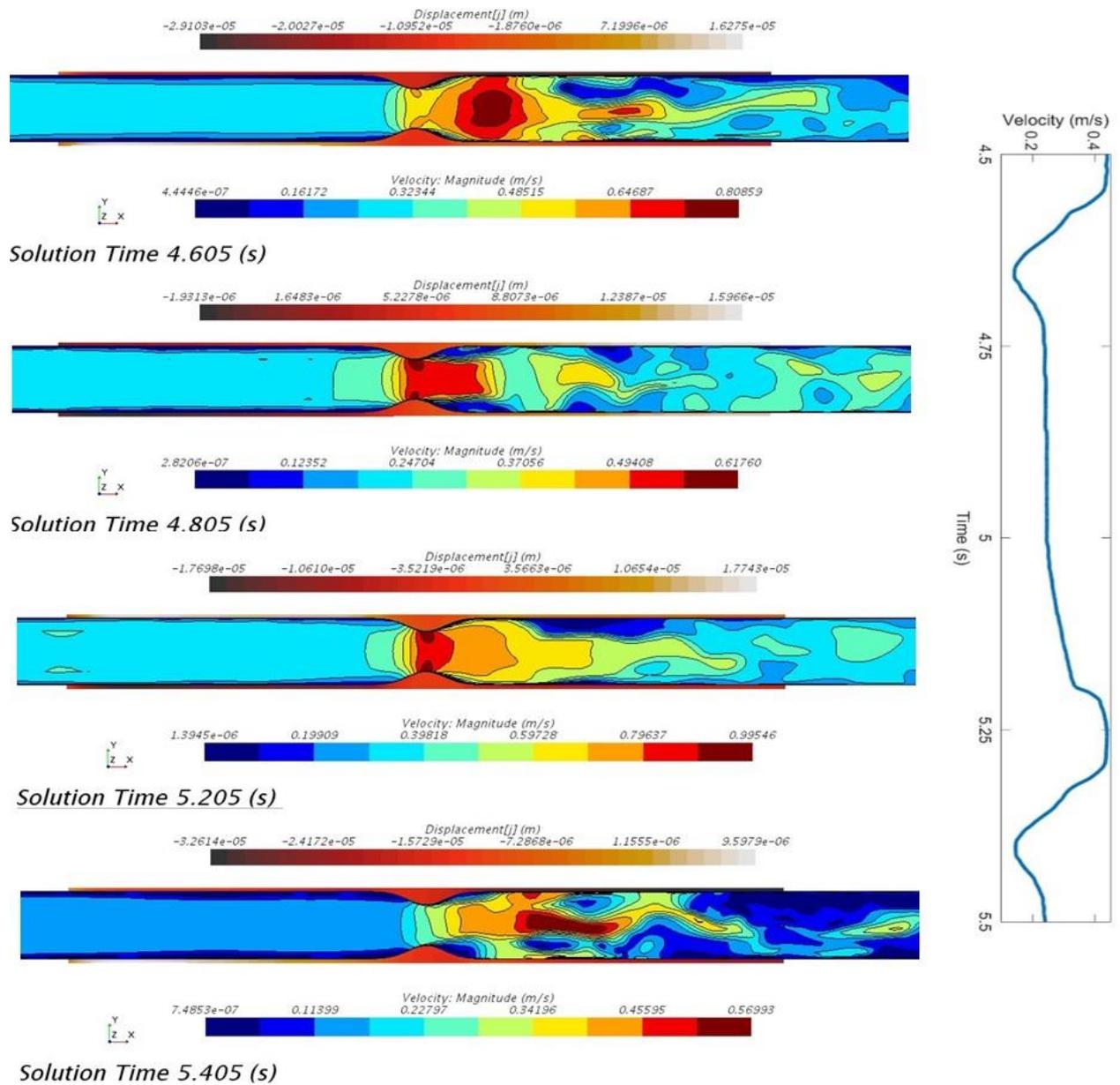

**Fig**. 7  Maps of velocity contours and wall displacements on a cross-section of the artery at different times during the cardiac cycle

## 4. CONCLUSION

This study incorporates two way coupled FSI modelling to study the propagation of wall displacement pulses in an artery wall caused by the pulsatile blood flow. Simulations were performed for a healthy artery and a stenosed artery to study the effects of geometrical changes due to the stenosis on pulse wave propagation. The simulation was validated by comparing the simulation PWV with the theoritical value for a healthy artery. Propagation of wall displacements were analysed using spatiotemporal maps and significant differences between the healthy and stenosed arteries were observed. Some propagating waves are damped due to the increase wall thickness in the stenosed region and some waves were generated after the stenosis possibly caused by the high veocity fluctuations induced by the instability and the breakdown of the high velcity jet generated at the stenosis. The results from this study may provide insights for future numerical and in vivo experimental studies to analyse pulse wave propagation.